\documentstyle[prb,aps,epsf]{revtex}
\begin{document}
\title{
Cyclotron effective masses in layered metals
}
\draft
\author{Jaime Merino\cite{email0} and Ross H. McKenzie\cite{email} }
\address{School of Physics, University of New
South Wales, Sydney 2052, Australia}
\date{\today}
\maketitle
\widetext

\begin{abstract}
Many layered metals such as quasi-two-dimensional organic
molecular crystals show properties consistent with a Fermi liquid
description at low temperatures.
 The effective masses extracted from the temperature
dependence of the magnetic oscillations
observed in these materials are in the range, $m^*_c/m_e \sim 1-7$,
suggesting that these systems are strongly correlated.
However, the ratio $m^*_c/m_e$ contains
both the renormalization due to the electron-electron
interaction and the periodic potential of the lattice.
We show that for {\it any} quasi-two-dimensional band structure,
the cyclotron mass is proportional to the density of states
at the Fermi energy. Due to Luttinger's theorem, this
result is also valid in the presence of interactions.
We then evaluate $m_c$ for several model band
structures for the $\beta$,  $\kappa$, and $\theta$ families of
(BEDT-TTF)$_2$X, where BEDT-TTF is
bis-(ethylenedithia-tetrathiafulvalene) 
and X is an anion. We find that for $\kappa$-(BEDT-TTF)$_2$X,
the cyclotron mass of the $\beta$-orbit,
 $m^{*\beta}_c$,
 is close to 2 $m^{*\alpha}_c$,
where $m^{*\alpha}_c$ is the effective mass of the $\alpha$-
orbit. This result is fairly insensitive to
the band structure details. For a wide range of materials we compare values
of the cyclotron mass deduced from band structure calculations
to values 
deduced from measurements of magnetic oscillations
and the specific heat coefficient $\gamma$.
\end{abstract}

\pacs{PACS numbers: 71.27.+a, 71.10.Fd}

\subsection{Introduction}
Quasi-two-dimesional metals such as the organic molecular
crystals based on the BEDT-TTF molecule [where BEDT-TTF is
bis-(ethylenedithia-tetrathiafulvalene)] 
and the layered
perovskites Sr$_2$RuO$_4$,
show properties which are consistent with a Fermi liquid description
at low temperatures.\cite{merino:99}
Although transport properties of these materials show unconventional
behaviour with temperature at high temperatures,
at low temperatures (below about 20 K in the organics)
the resistivity is quadratic with temperature,
the thermopower is linear in temperature and a Drude peak
is present in the optical conductivity.\cite{Ross:98} Furthermore, 
magnetic oscillations 
such as the de Haas - van Alphen effect is observed\cite{Wosnitza,Mackenzie:96}
suggesting the presence of a well defined Fermi surface 
and quasiparticle excitations described by
Fermi liquid theory.                  
In order to understand the role of
electron-electron interactions in
these materials it is then necessary to quantify the strength
of electron correlations and test how robust the Fermi liquid
description is.

Cyclotron effective masses for the quasiparticles can be 
obtained from fitting the observed temperature
 dependence of the amplitude  of magnetic
oscillations
to the Lifshitz-Kosevich form.
The amplitude at a temperature $T$ is proportional to 
\begin{equation}
R_T = {X \over \sinh X}
\ \ \ \ \ \ \ \ \ \ \ \ \ \ \  X = {2 \pi^2 k_B T  \over \hbar \omega_c^*}
\label{split1}
\end{equation}
where $\omega_c^* = e B/ m_c^*$ is the cyclotron frequency
and $m_c^*$ is the cyclotron effective mass, including many-body
effects.

Typical values obtained for the cyclotron mass
in these materials are in the range,
 $m^*_c/m_e \sim 1-7$, (where $m_e$ is the
free electron mass)
suggesting the possibility that many-body effects may
cause a significant enhancement of the
effective mass.
 However, knowing $m^*_c/m_e$ by itself is not sufficient
to determine the size      of many-body effects
due to electron-electron and electron-phonon interactions.
 First, it is necessary to compute
the cyclotron band mass, $m_c$, which takes into account the
fact that electrons are not free but are moving in the
presence of the periodic potential associated with the
crystal lattice.
Then, the ratio $m^*_c/m_c$ can be used to estimate the
importance of many-body effects.
 As we will
see, estimates of $m_c$ deduced from band structure calculations, can vary by
as much as a factor of three.

On the other hand, recent calculations
of the transport properties of strongly correlated systems
using dynamical mean-field theory \cite{merino:99} to solve the Hubbard 
model
on a frustrated hypercubic lattice, indicate that as the electronic
correlations become stronger  there is a clear crossover
from a Fermi liquid at low temperatures to
a ``bad metal'' with no quasiparticles at high temperatures.
However, such a crossover and the associated
signatures in transport properties
(e.g., a peak in the temperature dependence of the
thermopower and resistivity, and disappearance of the
Drude peak in the optical conductivity)
are only observed for sufficiently
large values of the ratio: $m^*_c/m_c \sim 3-4$.
For smaller values, transport properties resemble
the ones found in a nearly free-electron metal.
Since the transport properties of the organic metals do show the
signatures discussed above it is important
to have  accurate estimates of $m^*/m_c$ in order 
to check the consistency of describing them
as strongly correlated systems.

In this paper, we show that in a
quasi-two-dimensional Fermi liquid 
there is a simple relation between the cyclotron mass 
and the density of states at the Fermi surface.
This result, Eq. (\ref{cyclo}), 
holds for {\it any} dispersion relation for the 
quasiparticles.
Using this relation, we compute the ratio of
the cyclotron band masses associated with the $\alpha$ and $\beta$ orbits,
$m_c^{\beta}/m_c^{\alpha}$,
for a model band structure for $\kappa$-(BEDT-TTF)$_2$X.
The ratio is approximately 2, and varies by only about
ten per cent even when the band structure  parameters are
varied significantly.
This is in good agreement with values of
$m^{*\beta}/m^{*\alpha}$, 
deduced from magnetic oscillation experiments, suggesting that
the quasiparticle renormalisation factor
does not vary significantly between different parts
of the Fermi surface.

\subsection{Different effective masses}

We now briefly review several  of the effective masses
which can be defined for
electrons or quasiparticles with a general        
dispersion relation $\epsilon({\bf k})$.

{\it Band mass tensor.}
This is defined as\cite{Kittel}
\begin{equation}
m^b_{\nu \mu} \equiv \hbar^2 
 \left( {\partial^2 \epsilon({\bf k}) \over   \partial k_{\nu}
 \partial k_{\mu}}\right)^{-1} 
\label{mband}
\end{equation}
where $\nu$ and $\mu$ are Cartesian coordinates,
and gives information of the band dispersion for any 
value of the electronic momentum.                
In particular, from the band mass tensor, the band dispersion
of the electrons in {\it all} directions near the Fermi surface
can be reconstructed.

{\it Cyclotron mass.}
When a metal is in the presence of
an external magnetic field 
the electrons undergo periodic
orbits in both position and momentum space.
 The cyclotron frequency, $\omega_c$, associated
with the periodic motion along these orbits
on the Fermi surface is given by\cite{Ashcroft}
\begin{equation}
{1 \over \omega_c} =
{\hbar^2 \over  2 \pi e B }
\oint {dk \over
(\nabla \epsilon({\bf k}))_{\perp} }
\equiv
{ m_c \over e B}
\label{omegac}
\end{equation}
where $B$ is the strength of the magnetic field and
$ ( {\bf \nabla} \epsilon({\bf k}))_{\perp} $, is the gradient of the 
dispersion relation in the plane perpendicular to the field
and the line integral is around the periodic orbit on the Fermi surface. 
The last relation has been used to define a 
cyclotron effective mass, $m_c$.
Note that this effective mass involves an average of
the dispersion relation along the periodic orbit.
It determines the energy spacing of the
Landau levels and can be extracted from the
temperature dependence of the amplitude of
magnetic oscillations, as discussed above.

{\it Plasma frequencies.}
Reflectivity measurements can be used
to determine the plasma frequency associated
with collective oscillations of a charged Fermi liquid.
Polarized light can be used to determine the
anisotropy of these frequencies.
For light polarized with the electric field in the $\mu$ direction 
in a metal with Fermi energy $\epsilon_F$, the plasma frequency,
 $\omega_{p\mu}$, is given by,\cite{Kwak}  
\begin{equation}
\omega^2_{p\mu} = ({ e \over \pi \hbar} )^2 \int{  d^3 {\bf k} {\partial^2 \epsilon( {\bf k}) 
\over \partial k_\mu^2 } \theta(\epsilon_F - \epsilon_{\bf k})  }
\equiv {n e^2 \over m_{p \mu}}
\end{equation}
where the integral runs over the first
Brillouin zone and the last identity has been used to define
an effective mass, $m_{p \mu}$, when 
$n$ is the total number of charge carriers.
 The above expression
is derived from Lindhard's dielectric function.
 Note that
 in contrast with the cyclotron mass in Eq. (\ref{omegac}),
 which depends on electron states at the Fermi surface,
the plasma mass includes all the occupied states, 
and not only those which are close to the Fermi energy.
 This is because the  
plasma oscillation is a collective process
 in which all the electrons participate.

For a parabolic dispersion relation, $\epsilon( {\bf k}) = \hbar^2
k^2 /(2m_0)$, all of the effective masses defined above
will equal $m_0$. However, we stress that for a general dispersion
relation they will {\it not} be equal and
so caution is in order when trying to
compare effective masses extracted from different measurements.

\subsection{The cyclotron mass and the density of states}

 We know show how
for a quasi-two-dimensional metal,
the cyclotron mass defined 
by (\ref{omegac}) is simply related
to the density of states at the Fermi energy.
First, following Ashcroft and Mermin,\cite{Ashcroft} Eq. (\ref{omegac})
can be rearranged to give
\begin{equation}
m_c={\hbar^2  \over 2 \pi }  {{\partial A(\epsilon_F)}
\over{ \partial \epsilon_F }}
\label{band}
\end{equation}
where $A(\epsilon_F)$ is the area of the cross section of the Fermi
surface defined by the orbit described by an electron or hole,
in the presence of a the magnetic field, $B$.

For a quasi-two-dimensional system with only one band
that crosses the Fermi energy,
and a magnetic field perpendicular to the layers,
the area of the orbit is just the cross sectional
 area of the Fermi surface within a layer 
\begin{equation}
A(\epsilon_F)=4 \pi^2 \sum_{\bf k} 
\theta(\epsilon_F-\epsilon( {\bf k}))
\label{area}
\end{equation}
where {\bf k} is the two-dimensional wavevector within a layer.
Eq.(\ref{area}) is
just based on state counting and assumes that the
interlayer dispersion can be neglected.
Corrections due to a finite interlayer bandwidth
will be of order $t_c/\epsilon_F$ where $t_c$
is the interlayer hopping integral.
For typical organic metals this ratio is less than
0.01.\cite{Wosnitza,moses}
Taking the derivative of (\ref{area})
with respect to $\epsilon_F$ gives, for the
cyclotron mass
\begin{equation}
m_c=2 \pi \hbar^2 \rho_{\sigma}(\epsilon_F)
\label{cyclo}
\end{equation}
where $\rho_\sigma(\epsilon_F)$ is the density of states per spin
at the Fermi energy, $\rho_\sigma(\epsilon_F)=\sum_{\bf{k}}
\delta(\epsilon_F-\epsilon(\bf{k}))$. We stress that
this simple expression for the cyclotron band mass is only true for
quasi-two-dimensional metals.
In other cases, the reduction of the general expression (\ref{band})
to (\ref{cyclo}) cannot be done.
For example, for a three-dimensional
system the area associated to an electron or hole orbit is not defined
by Eq. (\ref{area}), and, therefore,
it is not possible to relate the cyclotron mass
to the density of states at the Fermi energy.
 The  result (\ref{cyclo}) was previously pointed out by
 Tamura {\it et al.}\cite{tamura}
 but its significance appears to have been completely overlooked.
We will show below that as a consequence of Luttinger's theorem
it is also true in the presence of interactions.

For more general situations where the Fermi surface of the metal
crosses several bands, the different cyclotron masses
can be expressed in terms
of the partial density of states associated with each of the
bands. As an example, we will compute the band
cyclotron masses for the $\alpha$ and
$\beta$ orbits in the $\kappa$-(BEDT-TTF)$_2$X family, for which
several bands are present.

\subsection{Model band structures}

\begin{figure}
\centerline{\epsfxsize=9cm \epsfbox{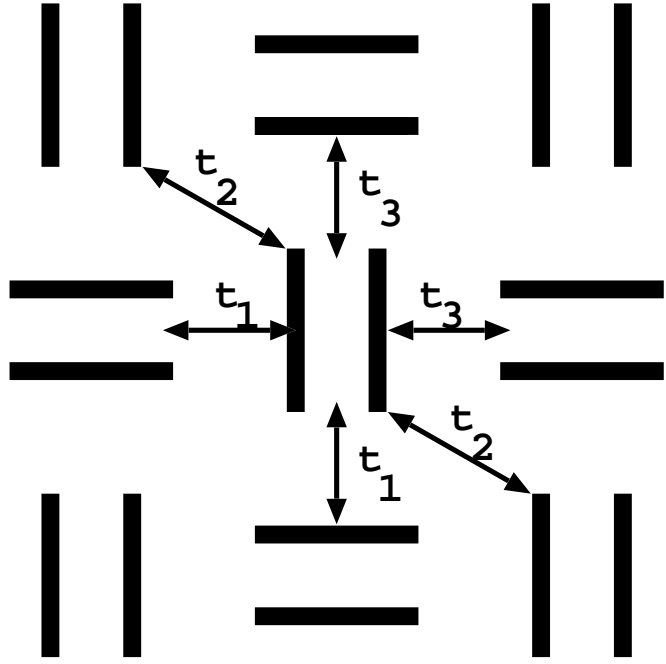}}
\caption{
 Stacking pattern of the BEDT-TTF
molecules within a layer of the  $\kappa$-(BEDT-TTF)$_2$X
family of organic metals. $t_1$, $t_2$ and $t_3$ denote hopping amplitudes
between {\it dimers} of molecules.
}
\label{stakappa}
\end{figure}

There are several approaches used for calculating the
band structure of layered materials. Semi-empirical
approaches such as the H\"uckel approximation use
parametrized tight-binding Hamiltonians with parameters
that are partially determined from experiment. In the case of
$\kappa$-(BEDT-TTF)$_2$X crystals, the effective tight-binding
Hamiltonian which is used to model the interaction between the
antibonding orbitals of BEDT-TTF dimers
 is\cite{Caulfield,visentini,fuku,Ross:98}
\begin{equation}
H=t_1 \sum_{ij}( c^\dagger_i c_j + h. c. )
+ t_3 \sum_{ik} ( c^\dagger_i c_k + h. c. )
+ t_2 \sum_{il} (c^\dagger_i c_l + h. c. )
\label{tight}
\end{equation}
where $c^{\dagger}_i$, creates an electron in the antibonding orbital
at site $i$ on a square lattice.
$t_1$ and $t_3$ are nearest-neighbour hoppings, and $t_2$ is
the next-nearest neighbour hopping amplitude along only {\it one}
diagonal. The $\kappa$-(BEDT-TTF)$_2$X materials  have
two dimers per unit cell and, because
$t_1$ and $t_3$ can be slightly different the
two dimers in each cell of the $\kappa$-(BEDT-TTF)$_2$X materials
are inequivalent. The relationship between the 
different hopping integrals and the geometrical
arrangement of the BEDT-TTF molecules
is shown in Fig. \ref{stakappa}.

\begin{figure}
\centerline{\epsfxsize=10cm \epsfbox{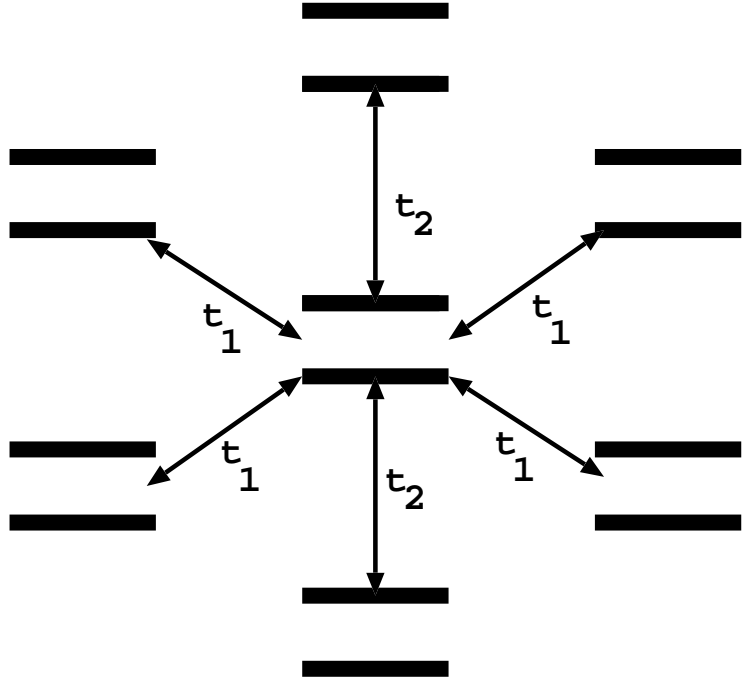}}
\caption{ 
 Stacking pattern of the BEDT-TTF
molecules within a layer of the  $\beta$-(BEDT-TTF)$_2$X
family of organic
superconductors. $t_1$ and $t_2$ denote hopping amplitudes
between {\it dimers} of molecules.
}
\label{stabeta}
\end{figure}

In Fig. \ref{stabeta} we show the stacking pattern for the 
$\beta$-(BEDT-TTF)$_2$X family.  In this case all the sites 
in the lattice are equivalent and there is only one dimer per unit cell.

\begin{figure}
\centerline{\epsfxsize=10cm \epsfbox{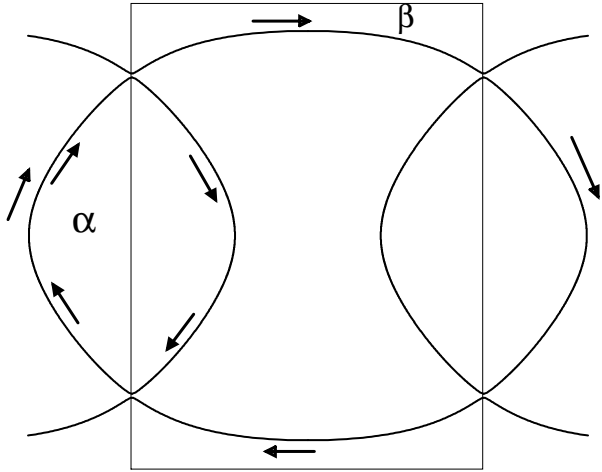}}
\caption{ $\alpha$ and $\beta$ orbits on the intralayer Fermi
surface in a $\kappa$-(BEDT-TTF)$_2$X material.
The arrows indicate the motion of holes on each
orbit in the presence of a magnetic field
perpendicular to the layers. The box determines the Brillouin zone boundary. 
There is a small energy gap at the boundary and the
$\beta$ orbit is only observed in magnetic oscillations
due to magnetic breakdown.
In the $\beta$-(BEDT-TTF)$_2$X family the
Brillouin zone is twice as large and there is no $\alpha$ orbit.}
\label{fermi}
\end{figure}

If we diagonalize the Hamiltonian (\ref{tight}), 
we obtain the two dispersion relations
\begin{equation}
\epsilon^{\pm}( {\bf k})= t_2 \cos(k_y) \pm   
\left( t_1^2 + t_3^2 + 2 
t_1 t_3 \cos(k_x) \right)^{1/2} \cos(k_y/2)
\label{kappadisp}
\end{equation}
The Fermi surface is shown in Fig. \ref{fermi}, for
$t_1-t_3=0.05$ and $t_2=t_1$.
The $\alpha$ orbit is associated with the hole pocket in the
Fermi surface and the unoccupied part of the lower
band, $\epsilon(\bf{k})$, while the $\beta$ orbit
(which occurs in large magnetic fields due to
magnetic breakdown)
contains parts from both the upper and
lower band dispersions and corresponds to the outer orbit
described with arrows in Fig. \ref{fermi}.

For the $\beta$-(BEDT-TTF)$_2$X materials, due to the columnar
stacking of the BEDT-TTF molecules and being $t_1 \sim t_2 \sim t_3$,
there is only one dimer of
BEDT-TTF molecules per site, and there is only one
half-filled band which cuts the Fermi energy and is described by
\begin{equation}
\epsilon( {\bf k}) = t_2 \cos(k_y) + 2 t_1 \cos(k_x/2) \cos(k_y/2)
\end{equation}

For the $\theta$-(BEDT-TTF)$_2$X materials, the geometrical arrangement
is similar to that for $\kappa$-(BEDT-TTF)$_2$X with each dimer
replaced by a single BEDT-TTF molecule.\cite{mori}
It is then described by the dispersion relation (\ref{kappadisp})
but the band is 3/4 filled.

We have evaluated the cyclotron band masses associated with the
different orbits described along the Fermi surface
for $\kappa$-(BEDT-TTF)$_2$X. The area associated with the
$\alpha$-orbit (see Fig. \ref{fermi}) is given by
\begin{equation}
A^{\alpha}(\epsilon_F)=4 \pi^2 \sum_{ \bf{k}}
(1- \theta(\epsilon_F-\epsilon^-(\bf{k})))
\end{equation}
and the cyclotron effective mass is, from Eq. (\ref{cyclo})
\begin{equation}
m^{\alpha}_c=2 \pi \hbar^2 \rho_{\sigma}^-(\epsilon_F)
\end{equation}
where $\rho_{\sigma}^-(\epsilon_F)$, is the density of states per unit cell
and spin associated with the $\epsilon^-(\bf{k})$ band.
Similarly, the area enclosed by the $\beta$-orbit is

\begin{equation}
A^{\beta}(\epsilon_F)=4 \pi^2  \sum_{ \bf k} (1-
\theta(\epsilon_F-\epsilon^+({\bf k} )))  + 4 \pi^2
\sum_{ { \bf k}} (1- \theta(\epsilon_F-\epsilon^-(\bf{k})) )
\end{equation}
and the cyclotron mass is proportional to the total density
of states:
\begin{equation}
m^{\beta}_c=2 \pi \hbar^2 \rho_{\sigma}(\epsilon_F)
\end{equation}
where $\rho_{\sigma}(\epsilon_F)$ is the total density of states
per unit cell and spin. Note that a minus sign comes in the above
expressions when we are considering the electron mass instead of the 
hole mass as $m_e=-m_h$, where $m_h$ is the hole mass.

For the dispersion (\ref{kappadisp}) with
$t_1=t_2=t_3=t$,
Ivanov,  Yakushi, and Ugolkova\cite{Ivanov:97} have
obtained analytical expressions for the
density of states projected onto the upper and
lower bands can be obtained. If all energies
are in units of $t$, the total density of states
per unit cell and spin is      
\begin{eqnarray}
\rho(-3/2 \le \epsilon \le -1 )& = & {2 \over  \pi^2 q \sqrt{\tau} } K({1 \over q})
\nonumber \\
\rho(-1 \le \epsilon \le 3 ) & = & {2 \over  \pi^2 \sqrt{\tau} } K(q)
\nonumber \\
\label{Ivan1}
\end{eqnarray}
and, for the partial density of states associated with the lower band 
\begin{eqnarray}
\rho^-({-3 \over 2} \le \epsilon \le -1 ) & = & {2 \over \pi^2 q \sqrt{\tau} } K({1 \over q})
\nonumber \\
\rho^-(-1 \le \epsilon \le 1 ) & = & { 2 \over \pi^2 \sqrt{\tau} } F( \arcsin
\left( {  {1 \over {2q} } } \sqrt{ (5- \tau^2) (\tau+1) \over 2}\right) ; q )
\nonumber \\
\label{Ivan2}
\end{eqnarray}
where $q=\sqrt{ 1 - (\tau-1)^3(\tau +3)/(16\tau) }$
with $\tau= \sqrt{2 \epsilon +3 }$. $K$ and $F$ are the complete elliptic
integral and the elliptic integral of the first kind,
respectively. 

From the above expressions and Eq. (\ref{cyclo}) we obtain the 
following cyclotron masses:
 $m^{\beta}_c/m_e=0.23/t$ and $m^{\alpha}_c/m_e=0.11/t$,
 with $t$ given in eV and we have used the intralayer
unit cell area of $A=104.\AA^2$.
This gives $m^{\beta}_c/m^{\alpha}_c=2$ and it turns out
that this ratio is relatively insensitive to
variations in the band structure parameters.
 We have relaxed the condition 
on the hopping integrals $t_1=t_2=t_3$, and, we have numerically 
evaluated the partial density of states instead of using eq.(\ref{Ivan1})
and eq.(\ref{Ivan2}). 
The ratio of the
cyclotron masses obtained from the effective dimer
model for fixed $t_1=t_3$ but different values of $t_2/t_1$ is,  
 $m_c^{\beta}/m_c^{\alpha}$ =
2.4, 2.2, and 2.0,
for $t_2/t_1=0.5, 0.7, 1.0$, respectively. 

In order to have a realistic description of the layered materials
we use the  hopping amplitudes obtained
from  quantum chemistry calculations using the H\"uckel approximation and,
in some cases, results obtained from first-principle calculations.
The hoppings of the effective dimer model, for which 
$t_1=t_3$ and $t_1 \ne t_2$, are given in Table \ref{table1}.        
A more detailed discussion of this model and the 
relationship between $t_1 $ and $ t_2$ and
the intermolecular hoppings calculated in the H\"uckel approximation
can be found in Reference \onlinecite{Ross:98}.
 A minor point is that if
we denote the Coulomb repulsion 
in each molecule by $U_0$, and the hopping amplitude between the
molecules within one dimer by $t_b$, for $U_0 >> 4 t_b$ (strongly correlated 
case),
the hopping amplitudes should be corrected by a factor of $1/\sqrt{2}$
with respect to the ones obtained from the H\"uckel calculation.
However, in the case $U_0 \sim 4 t_b$,
this factor is 0.92 and the effect of
correlations to the matrix elements is small.
Different calculations suggest that the ratio $U_0/4 t_b \sim 1$, 
so that in Table \ref{table1} we multiply all the bare hoppings by
0.92.

 In Table \ref{table1}, we also give the cyclotron masses obtained
from eq.(\ref{cyclo}), where the density of states has
been computed numerically
for the different hoppings. It can be seen that the
calculated cyclotron band masses are
sensitive to the parameters and the values deduced from the parameters
calculated by different groups for the same
material can vary significantly.                    
However, the calculated ratio, $m_c^{\beta}/m_c^{\alpha}$, is relatively 
insensitive to the parameters.

The band structures of the (BEDT-TTF)$_2$X family
have been calculated by several different techniques 
and some of the results for the density of states
at the Fermi energy are compared in Table \ref{table2}.
The H\"uckel method is the simplest and only considers
the $\pi$ orbitals and neglects all $\sigma$ orbitals.
The overlap integrals that are calculated are all
scaled by some empirical parameter and then
used as hopping integrals in a tight-binding band structure.
It is generally acknowledged that this method gives a good
qualitative description of electronic properties
(such as the symmetry and ordering of states)
but cannot give a quantitative description of
electronic properties.\cite{lowe}

The extended H\"uckel method\cite{Whangbo} treats both
$\pi$ and $\sigma$ orbitals.
Although it is more quantitatively reliable than
the H\"uckel approximation it still does not give
a completely quantitative description of organic molecules.
It has been used to calculate the band structure of
a wide range of organic metals by Whangbo and coworkers\cite{williams}.

The energy levels for a pair of BEDT-TTF dimers with
the same geometrical arrangement as in
$\kappa$-(BEDT-TTF)$_2$Cu[N(CN)$_2$]Br 
have been calculated by an        {\it ab initio}
method.     
The tight-binding parameters for a Hubbard model
for the dimers is then evaluated by fitting the
energy levels to the {\it ab initio} values.
The resulting parameters are similar to those obtained
by an extended H\"uckel calculation for the dimer pair.\cite{Fortunelli}
But the resulting density of states is more than twice
the results of extended H\"uckel for the solid.

The most reliable method of calculating band structures
is generally  considered to be {\it ab initio}
methods based on the local density approximation (LDA).
Nevertheless, different groups still often obtain
quite different results. For example,
values obtained for the density of states at the Fermi
energy in the fullerene metal, K$_3$C$_{60}$,
differ by as much as fifty per cent.\cite{Gunnarsson}
(Extended H\"uckel calculations do fall in this range).
Due to the large number of atoms in a unit cell
only a few {\it ab initio} calculations have been attempted for the
(BEDT-TTF)$_2$X materials.

Results for the density of states (and the corresponding
cyclotron masses) obtained using the three methods
are shown in Table \ref{table2}.
Note the large variation in results for each of the materials.
In particular, the H\"uckel method gives masses that
are two to five times larger than those obtained by
the other more sophisticated methods.

\subsection{The cyclotron mass in the presence of interactions}

The above treatment neglected the effect of
interactions between the electrons.
We now show that Eq. (\ref{cyclo}) has a natural
generalization in the case of a Fermi liquid.
The one-electron Green's function in a general interacting electron
system is
\begin{equation}
G({\bf k},\omega+ i \eta)=
 {1 \over \omega +i \eta - \epsilon({\bf k} ) - \Sigma( {\bf k},\omega) }
\end{equation}
in momentum space, where $\Sigma( {\bf k},\omega)$ is the electron self-energy.
In a Fermi liquid, near the quasiparticle poles, 
the Green's function can be rewritten as
\begin{equation}
G({\bf k},\omega)={ Z_{\bf k} \over \omega  - \tilde{\epsilon}({\bf k}) } 
\end{equation}
where $\tilde{\epsilon}({\bf k})$ is the quasiparticle energy and
$Z_{\bf k}= {1 \over 1- {\partial \Sigma ({\bf k},\omega) \over \partial \omega } |_{\omega=
\tilde{\epsilon}({\bf k}) } } $ is the residue at the quasiparticle pole.
Note that the above expression is true for
a Fermi liquid, and for electrons with momentum close to the Fermi surface
for which Im$\Sigma( {\bf k} \rightarrow {\bf k_F}
,\omega \rightarrow \epsilon_F) \rightarrow  0$. The spectral density is
then given by
\begin{equation}
A({\bf k},\omega)= -{1 \over \pi} {\rm Im} G( \omega + i \eta ) =
{\delta(\omega - \tilde{\epsilon}({\bf k}) ) \over  1- {\partial \Sigma ({\bf k},\omega) 
\over \partial \omega }|_{\omega= \tilde{\epsilon}({\bf k}) } }
\end{equation}
Thus, the quasiparticle density of states at the Fermi energy is
\begin{equation}
\tilde{\rho}(\tilde{\epsilon}_F) =
 \sum_{\bf k} \delta( \tilde{\epsilon}_F - \tilde{\epsilon}({\bf k} )  ) = 
 \sum_{\bf k} ( 1- {\partial \Sigma ({\bf k},\omega) \over \partial \omega }|_{\omega=
 \tilde{\epsilon}_F }  ) A({\bf k},\tilde{\epsilon}_F) 
\label{quasip}
\end{equation}
M\"uller-Hartman \cite{Muller} showed that, if the self energy is independent
of momentum, then at zero temperature,
 $\sum_{\bf k}A({\bf k},\tilde{\epsilon}_F)
=\rho(\epsilon_F)$, the non-interacting density of states at
the Fermi energy.
So in this case, $\tilde{\rho}(\tilde{\epsilon}_F) =
\rho(\epsilon_F)/Z$.
 Note that the quasiparticle
density of states 
is always enhanced because for a Fermi liquid ${\partial \Sigma ({\bf k},\omega) \over \partial \omega }|_{\omega= \tilde{\epsilon}_F }  < 0$.

Some time ago,
Luttinger \cite{Luttinger:61} showed  that in 
an interacting system with
Fermi liquid properties, the results of Lifshitz and Kosevich
still describe the de Haas van Alphen oscillations
provided that the relevant quasiparticle quantities are used.
Thus, (\ref{band}) is replaced by $m^*_c = {\partial \tilde{A}\over \partial
\tilde{\epsilon}_F}$ where a tilde denotes renormalised quantities.
In a quasi-two-dimensional Fermi liquid, 
the area enclosed by the
orbits of the quasiparticles is
\begin{equation}
\tilde{A}(\tilde{\epsilon}_F)=4 \pi^2 \sum_{\bf k} \theta(\tilde{\epsilon}_F
-\tilde{\epsilon}( {\bf k}))
\label{Aquasp}
\end{equation}
and so, we find that the cyclotron effective mass is
\begin{eqnarray}
m^*_c = 2 \pi \hbar^2 \sum_{\bf k } \delta(\tilde{\epsilon}_F-
\tilde{\epsilon}({\bf k}) )
= 2 \pi \hbar^2 \tilde{\rho}_{\sigma}(\tilde{\epsilon}_F)
\label{recyclo}
\end{eqnarray}

Again, equations (\ref{recyclo}) and (\ref{quasip}) show the cyclotron mass 
enhancement produced by the factor appearing in eq.(\ref{quasip}).
The same enhancement also appears in the specific heat
 coefficient.\cite{Luttinger:60}
A further simplification is obtained for the case of
a momentum independent self-energy, as then the
cyclotron effective masses reduce to
\begin{equation}
m^*_c=2 \pi^2 \hbar^2 \rho(\epsilon_F)/Z=m_c/Z
\end{equation}
where $Z$ is the quasiparticle weight, which, in
terms of the self-energy, is: $Z = ( 1- {\partial
\Sigma(\omega) \over \partial \omega}|_{\omega=\tilde{\epsilon}_F} )^{-1}$. In
this case, the ratios of the cyclotron effective
masses associated with the quasiparticles moving
along different orbits, $m^{*\beta}_c/m^{*\alpha}_c$, should be the 
same as the ratios associated with the non-interacting system,
$ m^{\beta}_c/m^{\alpha}_c$.

A partial test of the momentum independent self energy
is provided
 by comparing the
measured ratios of the renormalized cyclotron masses
 in different orbits with the
cyclotron band mass ratios.                                   
This is done in Table \ref{table1}.
The relative consistency between the observed values
of this ratio and the band structure values
suggests that if there are sizeable renormalizations
due to many-body effects then these renormalizations
are not significantly
different on the different parts of the Fermi surface.
However, this consistency is only a necessary condition
but not sufficient
for having a momentum independent self energy, as  
cyclotron masses include averages over the Fermi surface
and, therefore, cancellations of
contributions from different parts of the Fermi surface may occur.      

Furthermore, in Reference \onlinecite{Caulfield}
the effective masses for $\kappa$-(BEDT-TTF)$_2$Cu(SCN)$_2$
were measured as the pressure was increased from
1 bar to 20 kbar.
$m^{*\beta}_c/m_e$ decreased from 6.5 $\pm$ 0.1 at 1 bar
to 2.7 $\pm$ 0.1 at 16.3 kbar. 
$m^{*\alpha}_c/m_e$ decreased from 3.5 $\pm$ 0.1 at 1 bar
to 1.4 $\pm$ 0.1 at 16.3 kbar. 
However, the ratio $m^{*\beta}_c/m^{*\alpha}_c$
has a constant value of 1.9 within error.
           
\subsection{Specific heat}

Measurements of the electronic specific heat in 
the (BEDT-TTF)$_2$X crystals and
Sr$_2$RuO$_4$ show
a linear temperature dependence at low temperatures,
consistent with a Fermi liquid description.
The corresponding specific heat coefficient $\gamma$
 is given in Table \ref{table3}
for some of these materials. This coefficient is
related to  
 the quasiparticle density of states at the Fermi energy,
$\tilde{\rho}(\tilde{\epsilon}_F)$, (see Eq. (\ref{quasip})) by 
\begin{equation}
\gamma= { 2 \pi^2 k_B^2\over 3 } \tilde{\rho} (\tilde{\epsilon}_F)
\label{gamma}
\end{equation}

Since the quasiparticle density of states is
also related to the cyclotron effective mass  by 
Eq. (\ref{recyclo}) the measured 
specific heat coefficient can be used
to calculate a corresponding cyclotron effective mass.
This has been done in Table \ref{table3}
for a range of organic materials.
The values obtained for $m^{*\beta}_c/m_c$ from specific heat measurements
agree for $\kappa$-(BEDT-TTF)$_2$I$_3$ and $\beta$-(BEDT-TTF)$_2$I$_3$
but not for the materials with copper in the anion.
Since this comparison does provide a quantitative test
of a Fermi liquid description further careful
measurements are justified, particularly on
a wider range of materials.

Such a comparison was also done recently
for Sr$_2$RuO$_4$ in Reference \onlinecite{Mackenzie:96},
where relation (\ref{recyclo}) was implicitly assumed,
presumably based on its validity for a parabolic
dispersion relation.
Our work provides a rigorous justification for this comparison.
In Sr$_2$RuO$_4$ there are three distinct Fermi surfaces
and the associated cyclotron masses deduced from
de Haas van Alphen oscillations were 
$m^{*}_c/m_e$=3.4, 7.5 and 14.6 for the $\alpha$, $\beta$ and $\gamma$ orbits, 
respectively.\cite{update}From the above discussion, it follows that the 
specific heat coefficient of Sr$_2$RuO$_4$ is related to the effective
masses by
\begin{equation}
\gamma ={\pi k_B^2 \over 3  \hbar^2} (m^{*\alpha}_c + m^{* \beta}_c +
m^{* \gamma}_c)
\label{gamma*}
\end{equation}
which comes from the fact that the total density of states is
just the sum of the density of states of the different
Fermi surfaces.
Evaluating  (\ref{gamma*}) 
 we obtain a specific heat coefficient of 36.7 mJ/(K$^2$ mol),
which agrees    with the measured value\cite{Mackenzie:98}
of 37.4 mJ/(K$^2$ mol).

\subsection{Conclusions}

We now summarize our results and their implications.
First, it was shown that in a quasi-two-dimensional metal
in which the dispersion perpendicular to the layers
can be neglected, the
cyclotron effective  mass for a particular orbit
in a general band structure
is simply related to the density of states at the
Fermi energy associated with the relevant band.
Second, it was shown that, due to Luttinger's
results for a Fermi liquid, a similar
relationship holds in the presence of interactions.

These results have a number of general applications
to layered metals which have Fermi liquid properties
at low temperatures.

(i) In order to evaluate the effective mass from
band structure it is not necessary to numerically
evaluate the derivative in (\ref{band}), as has been done previously
by a number of authors.
Instead (\ref{cyclo}) can be used together with the density of states
at the Fermi energy. This eliminates the need to perform the cumbersome
task of repeating the band structure calculations for many different
Fermi energies.

(ii) We found that for model band structures describing
the family $\kappa$-(BEDT-TTF)$_2$X,
the ratio of the effective mass for 
the $\beta$-orbit
to the mass for the $\alpha$ orbit is fairly insensitive to the
details of the band structure, having  a value close to two.

(iii) Our results imply that a quantitative test
of the Fermi liquid description of a layered metal
is to compare measurements of the cyclotron
effective mass to the linear coefficient in the 
specific heat.

(iv) The agreement between 
the ratio of the different measured cyclotron masses
and the ratio calculated from band structure, 
suggests a momentum independent self-energy, although
other experimental probes such as polarized Raman scattering,
photoemission spectra, and angular dependent magnetoresistance oscillations
are needed before making any definitive conclusion.

Based on comparison with a wide range of materials we conclude
the following.
First, the effective masses deduced from magnetic oscillations
and specific heat are consistent for 
Sr$_2$RuO$_4$ and for two out of four of the organic
materials considered.
For three out of four of the organic materials for which
data is avalailable the measured ratio
 $m^{*\beta}_c/m^{*\alpha}_c$ is consistent with
the band structure ratio $m^{\beta}_c/m^{\alpha}_c$.
Furthermore, for 
the $\kappa$-(BEDT-TTF)$_2$ Cu(NCS)$_2$
this ratio does not change under pressure while the
individual effective masses decrease by a factor of 2.5.
This suggests that the self energy does not vary significantly
over the different parts of the Fermi surface.
We also note that
the significant variation of the effective masses
with pressure cannot be explained 
in terms of band structure; it predicts a small
variation with pressure.

A comparison of the results of band
structure calculations using a range of methods
found that they produced a large range
in values for the density of states (and thus
the effective masses).
The H\"uckel method has
often been used to estimate the hopping
integrals in tight binding band structures
(as in Table \ref{table1}). It is less sophisticated
than the extended H\"uckel method which in turn is
less sophisticated than {\it ab initio} methods based on the
local density approximation.
We suggest that the H\"uckel method is producing
hopping integrals which are too small by a factor of
two to four.
The best strategy to evaluate these
integrals would be to  fit a LDA band structure
to a tight binding dispersion, such as (\ref{kappadisp}).
Such an approach was recently taken for Sr$_2$RuO$_4$.\cite{mazin}

We now come back to the central question of this paper:
 are the layered metals we
have considered strongly correlated?
A definitive answer is not possible because of the
large variation in values 
for the band cyclotron masses that have been
calculated by different band structure methods.
However, we suggest that due to their greater sophistication,
the local
density approximation and extended H\"uckel approximation
calculations give the most reliable values.
We suggest that the appropriate values for the band
cyclotron masses are those calculated by the local
density approximation and extended H\"uckel approximation.
The mass ratios given in Table \ref{table2} then
imply that $m^{*\beta}_c/m_c \sim 2.5-4$, suggesting appreciable quasiparticle
renormalization due to many-body effects. 
This is consistent with the strong temperature
dependence of the transport properties,
discussed in detail in Reference \onlinecite{merino:99}.

\newpage

\vskip0.1in
\begin{table}
\caption{Cyclotron effective masses
predicted by tight-binding band structures with values
of the hopping integrals given by different H\"uckel calculations.
EHA denotes the extended H\"uckel approximation. 
These masses are compared to values deduced from magnetic oscillation
experiments.
The cyclotron masses are obtained from Eq. (\ref{cyclo})
with the density of states computed from the tight-binding Hamiltonian
(\ref{tight}) for  the given values of the hopping integrals
$t_1$ and $t_2$ and with $t_1=t_3$.
Note that the ratio of the masses for the $\beta$ and $\alpha$
orbits in Figure \ref{fermi} depends weakly on the band structure 
parameters. Except for the second line, all the results
are for ambient pressure.}
\vskip0.1in
\label{table1}
\begin{tabular}{lddddddd}
& $t_1$ (meV) &  $t_2$ (meV) & Ref. & $m^{\beta}_c/m_e$ & 
$m^{*\beta}_c/m_e$ (expt) & $m_c^{\beta}/m_c^{\alpha}$ & $m_c^{*\beta}/m_c^{*\alpha}$(expt) \\
\hline
$\kappa$-(BEDT-TTF)$_2$Cu(SCN)$_2$  & 31.3 & 23.0 & \onlinecite{Mariam} & 8.5 
& 6.5 \cite{Caulfield} & 2.2 & 1.9\cite{Caulfield}  \\
 \ \ \ \ \ \ 7.4 kbar           & 40.5 & 24.8  & \onlinecite{Mariam} & 7.4
 & $\sim$ 3.5\cite{caulfield2} & 2.3 & $\sim$ 2.0 \\
$\kappa$-(BEDT-TTF)$_2$Cu[N(CN)$_2$]Br & 61.7& 32.7 &\onlinecite{Fortunelli} & 4.6 & 6.4 \cite{Weiss:97} & 2.35 & ? \\ \ \ \ \  
 & 62.1& 42.3 & \onlinecite{Komatsu} & 4.4 & 6.4 \cite{Weiss:97} & 2.2 & ? \\
\ \ \ \ ab initio  & 78.2 & 39.0 & \onlinecite{Fortunelli} & 3.8 & 6.4\cite{Weiss:97} & 2.35 & ?  \\
$\kappa$-(BEDT-TTF)$_2$Cu$_2$(CN)$_3$ & 50.1 & 53.0&\onlinecite{Komatsu} & 4.6 & 4.0\cite{Ohmichi} & 2.0 & ? \\
$\kappa$-(BEDT-TTF)$_2$I$_3$ & 70.0 & 40.5 & \onlinecite{Komatsu} &  4.1 & 3.9 \cite{Wosnitza} & 2.3&
2.0 \\
 & 54.0 & 34.0 & \onlinecite{tamura:91} &  5.3 & 3.9 \cite{Wosnitza} & 2.3&
2.0\\
$\kappa$-(BETS)$_2$GaCl$_4$ & ?  & ? &  & ? & 5.3\cite{pesotskii} & ? & 1.6 \\
$\kappa$-(BETS)$_2$C(CN)$_3 $ & EHA  & ? & \onlinecite{canadell}
 & 1.2 & 3.3\cite{canadell} & ? & 1.9 \\
$\theta$-(BEDT-TTF)$_2$ I$_3 $ & 42.0 & 64.0 & \onlinecite{Kobayashi} & 2.2 & 3.6 \cite{tamura} & 2.6 & 1.8 \\
$\beta$-(BEDT-TTF)$_2$I$_3$ & 60.0 & 42.0 & \onlinecite{Mori} & 4.3 & 4.2 \cite{Wosnitza:97} & - &  -  \\
\end{tabular}
\end{table}
\vskip0.1in
\begin{table}
\caption{Comparison of 
the density of states at the Fermi energy 
(and the associated effective masses) which is obtained from 
different methods of calculating band structure.
 LDA denotes  {\it ab initio} calculations using the
local density approximation. EHA denotes
the extended H\"uckel approximation and
HA denotes values from Table \ref{table1} ,
based on the H\"uckel approximation.    
 The cyclotron masses are calculated
from the density of states using Eq. (\ref{cyclo}) .
The density of states
$\rho(\epsilon_F)$ is given in units of states per
 unit cell per spin per eV.        
Note that the H\"uckel method gives
effective masses that are two to five times larger than the
other more sophisticated methods.}

\vskip0.1in
\label{table2}
\begin{tabular}{lddddddd}
& LDA & & EHA& & HA& & Expt. \\
& $\rho(\epsilon_F)$ &$m^{\beta}_c/m_e$ & $\rho(\epsilon_F)$ &
$m^{\beta}_c/m_e$ & $\rho(\epsilon_F)$ & $m^{\beta}_c/m_e$  & 
$m^{*\beta}_c/m_e$\\
\hline
$\kappa$-(BEDT-TTF)$_2$Cu(SCN)$_2$ & 6.4 \cite{Xu:95} & 2.6  & 4.2 \cite{Haddon} 
& 1.7 & 21.2 &  8.5 & 6.5  \\
$\kappa$-(BEDT-TTF)$_2$Cu[N(CN)$_2$]Br & 6.9 \cite{Ching:97}
&  2.7 & 4.4 \cite{Haddon} & 1.7 & 11.7 & 4.6 & 6.4   \\
$\beta$-(BEDT-TTF)$_2$I$_3$ & ?  &&3.1 \cite{Haddon} & 1.6 & 5.6  & 4.3
& 4.2\\
\end{tabular}
\end{table}

\begin{table}
\caption{Comparison of the cyclotron effective masses, $m^{*\beta}$, deduced 
from the measurements of magnetic       oscillations associated
with the $\beta$ orbit and
the masses deduced from the linear specific heat coefficient $\gamma$
and equations (\protect\ref{cyclo}) and (\protect\ref{gamma}) .
$A$ is the area of the unit cell within a layer and $m_e$ is
the free electron mass.
}
\vskip0.1in
\label{table3}
\begin{tabular}{ldddd}
 & A($\AA^2$) & $m^{*\beta}/m_e$ & $\gamma$(mJ/(K$^2$ mol) ) 
& $m^*/m_e (\gamma)$ \\
\hline
$\kappa$-(BEDT-TTF)$_2$Cu(SCN)$_2$ & 104.0 &  6.5  $\pm$ 0.1
\cite{Caulfield} & 25 $\pm$ 3\cite{Andraka:89} &  4.4 $\pm$ 0.5  \\
$\kappa$-(BEDT-TTF)$_2$Cu[N(CN)$_2$]Br & 108.6 & 5.4\cite{Mielke:97},6.4
\cite{Weiss:97} & 22 $\pm$ 3 \cite{Andraka:91},
25 $\pm$ 2\cite{elsinger}  & 4 $\pm$ 1 \\
$\kappa$-(BEDT-TTF)$_2$I$_3$ & 103.0 &3.9\cite{Wosnitza}
 & 19 $\pm$ 1.5 \cite{Wosnitza:96a}& 3.4 $\pm$ 0.3\\
$\beta$-(BEDT-TTF)$_2$I$_3$ & 56.3  &  4.2 $\pm$ 0.2 \cite{Wosnitza:97}  & 
 24 $\pm$ 3 \cite{Stewart:86} & 3.9 $\pm$ 0.5 \\ 
\end{tabular}
\end{table}

\acknowledgements
We thank R. McKinnon, N. Harrison, J. S. Brooks, J. Wosnitza, and E. Canadell
for helpful discussions.
This work was supported by the Australian Research Council.


\begin{references}
\bibitem{caulfield2} This value was estimated
by interpolating values given
in Ref. \onlinecite{Caulfield} for different pressures.
\end{references}
\end{document}